\documentclass[oldversion]{aa}

\usepackage{amsmath,amssymb,graphicx,mathptmx,natbib,ifpdf}

\authorrunning{E. Puchwein \& M. Bartelmann}
\titlerunning{Probing the dynamical state of galaxy clusters}

\begin{document}

\title{Probing the dynamical state of galaxy clusters}
\author{Ewald Puchwein and Matthias Bartelmann
    \institute{Zentrum f\"ur Astronomie der Universit\"at
    Heidelberg, ITA, Albert-\"Uberle-Str.~2, 69120 Heidelberg,
    Germany}}

\date{\emph{Astronomy \& Astrophysics, submitted}}

\abstract{We show how hydrostatic equilibrium in galaxy clusters can be quantitatively probed combining X-ray, SZ, and gravitational-lensing data. Our previously published method for recovering three-dimensional cluster gas distributions avoids the assumption of hydrostatic equilibrium. Independent reconstructions of cumulative total-mass profiles can then be obtained from the gas distribution, assuming hydrostatic equilibrium, and from gravitational lensing, neglecting it. Hydrostatic equilibrium can then be quantified comparing the two. We describe this procedure in detail and show that it performs well on progressively realistic synthetic data. An application to a cluster merger demonstrates how hydrostatic equilibrium is violated and restored as the merger proceeds.}


\maketitle

\section{Introduction}

Numerous observations show that galaxy clusters frequently exhibit irregular shapes and violent dynamics. On the theoretical front, simulations indicate that cluster-sized dark matter halos are often well described as triaxial ellipsoids, but not as spheres \citep{JI02.1}. Nevertheless, clusters are often interpreted as spherically-symmetric objects in hydrostatic equilibrium, which is a potential source of error. For example, \cite{HA06.1} show that intrinsic variations in clusters limit the accuracy of cluster gas mass estimates to about 10\% when using such simple assumptions.

Several authors tried to relax the restricting assumption of spherical symmetry and aimed at a joint analysis of different types of cluster data.  \cite{ZA98.1} suggested to base the reconstruction of axisymmetric, three-dimensional gravitational cluster potentials on the Fourier slice theorem, extrapolating Fourier modes into the ``cone of ignorance''. They applied their technique to simulated data and showed that it performs well \citep{ZA01.1}. \cite{DO01.3} followed a perturbative approach, and \cite{LE04.1} proposed to adapt parameters of triaxial halo models, all by combining different data sets such as X-ray, (thermal) Sunyaev-Zel'dovich (SZ) and gravitational-lensing maps. A similar method was applied to data by \cite{DE05.1}.

An alternative approach based on the iterative Richardson-Lucy deconvolution was suggested by \cite{RE00.1} and \cite{RE01.1}. It aims at the gravitational potential, assumes only axial symmetry of the main cluster body, avoids extrapolations in Fourier space, and can easily be extended to include additional data sets. In \cite{PU06.1} (hereafter Paper I) we developed the latter algorithm further. However, instead of aiming at the gravitational potential, which would require us to assume a relation between the gas distribution and the gravitational field, we proposed methods to reconstruct the three-dimensional cluster gas density and temperature distribution from X-ray and thermal SZ effect observations. These methods do not require any equilibrium assumption other than local thermal equilibrium and again assume only axial symmetry with respect to an arbitrarily inclined axis. Using synthetic observations of analytically modelled and numerically simulated galaxy clusters we showed that these reconstruction methods perform very well, even in the presence of observational noise, deviations from axial symmetry and cluster substructure.

In this work we use this gas reconstruction algorithm together with novel methods to reconstruct the three-dimensional gravitational potential from lensing data in order to probe hydrostatic equilibrium in galaxy clusters and to quantify the accuracy of mass estimates based on the assumption of hydrostatic equilibrium. We will also introduce methods to find three-dimensional reconstructions of the gravitational potential and of the mass profiles of relaxed galaxy clusters from X-ray and thermal SZ observations alone. All these methods are tested with synthetic observations of analytically modelled and numerically simulated galaxy clusters. Mass estimates based on the gas reconstruction and the assumption of hydrostatic equilibrium are compared to lensing mass estimates and to the original analytic or simulated masses.

\section{Three-dimensional cluster reconstruction techniques}
\label{sec:techniques}

\subsection{Cluster gas reconstructions from X-ray and SZ data}
\label{sec:gas_reconstructions}

In Paper I a novel technique to reconstruct the intra-cluster medium in three dimensions by a combined analysis of X-ray and thermal Sunyaev-Zel'dovich effect observations was introduced (hereafter also called XSZ reconstructions). It assumes only axial symmetry of the cluster halo with respect to an arbitrarily inclined axis and does not require any equilibrium assumption other than local thermal equilibrium. The iterative method is based on Richardson-Lucy deconvolution \citep{LU74.1,LU94.1} and is a generalisation of employing it to reconstructions of three-dimensional axisymmetric quantities form their projection along the line-of-sight \citep{BI90.1}.  Due to the assumed axial symmetry the reconstructed gas densities and temperatures depend only on the distance $R$ from the symmetry axis and the coordinate $Z$ along the axis.
 
In this study we employ the same three-dimensional gas reconstruction method as in Paper I except for one modification. We now use a more realistic model for the X-ray emission of the intra-cluster medium that also includes line emission. The synthetic X-ray observations of the reconstructed cluster halo that are performed during the iterative deprojection (see Paper I) are now calculated with the MEKAL emission model (see \cite{LI95.1,KA93.1}) and the WABS model for galactic absorption \citep{MO83.1}. More precisely we use the X-ray spectral fitting software package XSPEC \citep{AR96.1} to create a table of the cooling function with the models mentioned above, assuming a constant metallicity of 0.3 times the Solar value and an equivalent hydrogen column density of $5\times10^{20} \, \mathrm{atoms} \; \mathrm{cm}^{-2}$. We then use this table to produce the synthetic X-ray maps needed during the reconstruction and we also employed it to create the synthetic observations of analytically modelled and numerically simulated clusters that are discussed in sections \ref{sec:analytic} and \ref{sec:numerical}. Note that it is not necessary to change the equations for the iterative corrections of the gas density and temperature (Eq. (22) and (23) in Paper I), which were derived considering only thermal bremsstrahlung, because small errors introduced by using these equations are corrected in subsequent iteration steps. It also does not significantly affect the number of iterations needed to achieve a good reconstruction.

\subsection{Reconstructions of the gravitational potential and the total mass distribution from X-ray and SZ data}
\label{sec:x-ray_sz_pot_reconstruction}

We want to use these three-dimensional cluster gas reconstructions to find the gravitational potentials and total mass distributions of relaxed galaxy clusters and to probe the dynamical state of potentially unrelaxed clusters by comparing such gas reconstructions to an analysis of lensing data.

To find the gravitational potential of a cluster from the distribution of the cluster gas we assume that the gas is in hydrostatic equilibrium. Then the gas density $\rho$, the gas pressure $p$ and the gravitational potential $\phi$ satisfy 
\begin{equation}
  \vec{\nabla} \phi = -\frac{\vec{\nabla} p}{\rho}.
  \label{eq:equilibrium_condition}
\end{equation}	
In principle this equation can be used to find the gravitational potential of relaxed clusters from three-dimensional reconstructions of their intra-cluster medium. However due to deviations from hydrostatic equilibrium, and the presence of observational noise and cluster substructure violating axial symmetry, the curl of $-\vec{\nabla} p/\rho$ will not vanish exactly for the reconstructed gas distributions. Thus one cannot obtain a unique solution for $\phi$ directly from Eq. (\ref{eq:equilibrium_condition}).

To get a unique solution we first derive $-\vec{\nabla} p/\rho$ on the grid in $R$ and $Z$ space on which the gas reconstruction was calculated (see also Paper I). Then we aim to determine the potential $\phi$ for which $\vec{\nabla} \phi$  is closest to $-\vec{\nabla} p/\rho$. We do that by finding the values of the potential $\phi$ at all grid points which minimise the deviation
\begin{equation}
  \sum_{\mathrm{neighbours}\,\,i,j} \big(\phi_j - \phi_i+\frac{p_j-p_i}{\frac{1}{2}(\rho_j+\rho_i)}\big)^2,
  \label{eq:field_error}
\end{equation}
between these two vector fields. Here $p_i,p_j$ are the gas pressures, $\rho_i,\rho_j$ the gas densities and $\phi_i,\phi_j$ the gravitational potentials at the $R$ and $Z$ coordinates of grid points $i$ and $j$. The sum extends only over such pairs of grid points $i$ and $j$ that are nearest neighbours. Conjugate gradient minimisation starting with a guess $\phi_i=0$ is used to find the solution for the $\phi_i$.

However, to reduce noise in the potential it turned out to be favourable to add a penalty function to (\ref{eq:field_error}) that requires the second derivatives of the potential to be small. We also multiply each term in the sum  in Eq. (\ref{eq:field_error}) and the penalty function by a weight factor. So the function we end up minimising is,
\begin{align}
  & \sum_{\mathrm{neighbours}\,\,i,j} w(r_{i,j}) \big(\phi_j - \phi_i+\frac{p_j-p_i}{\frac{1}{2}(\rho_j+\rho_i)}\big)^2 + \nonumber \\
  & \quad \quad w_p \sum_{i} \big((\phi_{iR_>} + \phi_{iR_<}- 2\phi_i)^2 +  (\phi_{iZ_>} + \phi_{iZ_<}- 2\phi_i)^2\big),
  \label{eq:min_function}
\end{align}
where $r_{i,j}$ is the distance from the cluster centre to the midpoint of the line connecting grid points $i$ and $j$. $iR_>,iR_<$ and $iZ_>,iZ_<$ are the indices of the neighbouring grid points of point $i$ in the $R$ and $Z$ directions respectively. The weighting function $w(r)$ is chosen equal to one in the central region of the cluster, for $r< 0.3l$, then it smoothly goes to zero, and vanishes for $r>0.4l$, close to the perimeter of the box with side length $l$ that is used for the gas reconstruction (see Paper I). This is necessary because there are significant artefacts in the gas reconstruction close to the perimeter (see also Paper I). When using the potential reconstruction algorithm proposed here they would have a non-local effect on the potential reconstruction and would thereby reduce its quality also near the cluster centre. The weight factor $w_p = 3$ for the penalty function was chosen by trial and error and proved to be effective.

One can then use the reconstructed three-dimensional gravitational potential to find the total mass distribution of the galaxy cluster. 

A simpler alternative way to get mass estimates from the gas reconstruction and the assumption of hydrostatic equilibrium, that does not require us to reconstruct the gravitational potential, is to apply Gauss's law to the gravitational field and use Eq. (\ref{eq:equilibrium_condition}) to express the gravitational field as $\vec{\nabla} p/\rho$. This allows us to define a cumulative mass $M_{<r,\,\mathrm{XSZ}}$ as a function of radius $r$ from the cluster centre by
\begin{equation}
  M_{<r,\,\mathrm{XSZ}} \equiv \frac{1}{4\pi G} \int -\frac{\vec{\nabla} p}{\rho} d\vec{A},
  \label{eq:cum_mass_gas}
\end{equation}
where $G$ is Newton's constant and the integral extends over the surface of a sphere with radius $r$ around the cluster centre. The numerical evaluation of the integral is done using 128 sampling points which are equally spaced in the angular coordinate $\theta \equiv \arctan(Z/R)$. For each point the component of $\vec{\nabla} p/\rho$ perpendicular to the surface is calculated from the gas reconstruction and multiplied by the area of the corresponding ring.

\subsection{Reconstructions of the gravitational potential and the total mass distribution from lensing data}

Lensing observations allow reconstructions of the lensing potential (see e.g. \cite{CA05.1}), which is simply the suitably rescaled projection of the lensÕ gravitational potential along the line-of-sight. Once the lensing potential is found, Richardson-Lucy deconvolution can be applied to deproject it in order to obtain the three-dimensional gravitational potential. Again axial symmetry with respect to an arbitrarily inclined axis needs to be assumed. 

We employ the deprojection algorithm discussed in sections 2.1 and 2.2 of Paper I to obtain such three-dimensional reconstructions of the gravitational potential.

In Paper I the optimal number of iterations for three-dimensional gas reconstructions was studied. Richardson-Lucy deconvolution reproduces large scale structure quickly, while it converges slowly to small scale structure. It turned out that for gas reconstructions based on X-ray and SZ data it is best to use about five iterations. For a smaller number of iterations the cluster structure is not recovered sufficiently well, while for a larger number of iterations the reconstruction algorithm tries to reproduce small-scale observational noise which can reduce the reconstruction quality again. However as the lensing potential is a much smoother quantity than the X-ray surface brightness or the SZ temperature decrement it is favourable to use a larger number of iterations for deprojections of the lensing potential. 

However even when using a large number of iterations, problems with the gravitational potential reconstruction arise for small inclination angles $i$ between the line-of-sight and the symmetry axis, because then the assumption of axial symmetry contains least information (see also Paper I) and a reconstruction that, compared to the original halo, is stretched along the symmetry axis can still reproduce the lensing observations rather well. For a cluster with a roughly spherical gravitational potential and for a small inclination angle one gets too large correction factors close to the symmetry axis during the first few iteration steps when starting from a flat guess and thus the reconstruction after a few iterations is overly extended along that axis. As the power to determine the halo elongation along the symmetry axis is limited for small inclination angles the reconstruction algorithm takes very long to recover from this. To avoid this problem, it is thus favourable to start with a guess that has already more or less the right shape. We can get such a guess by doing a gravitational potential reconstruction from a flat guess with a small number of iterations and by then making the obtained potential spherically symmetric while preserving its profile. This spherically symmetrised potential can then be used as a first guess for the actual reconstruction with a larger number of iterations. 10 iterations were used to produce spherically symmetrised guesses for reconstructions from synthetic lensing data in sections \ref{sec:analytic} and \ref{sec:numerical}. The actual reconstructions use 30 iterations and start either from such a spherically symmetrised guess or from a flat guess as specified there.

The lensing three-dimensional gravitational potential reconstructions can then be used to find the total mass distribution and can be compared to reconstructions from X-ray and SZ data. In order to have a quantity that can be directly compared to $M_{<r,\,\mathrm{XSZ}}$ we define in analogy to Eq. (\ref{eq:cum_mass_gas}) a lensing cumulative mass
\begin{equation}
  M_{<r,\,\mathrm{lensing}}  \equiv \frac{1}{4\pi G } \int \vec{\nabla} \phi d\vec{A},
  \label{eq:cum_mass_lensing}
\end{equation}
where $\phi$ is the three-dimensional gravitational potential obtained by deprojecting the lensing potential. The numerical evaluation of the integral is done in the same way as for $M_{<r,\,\mathrm{XSZ}}$.

\subsection{Probing hydrostatic equilibrium}

Hydrostatic equilibrium in galaxy clusters can be probed by comparing cluster reconstructions based on X-ray and SZ data to lensing reconstructions. In principle this could be done by comparing the gravitational potential obtained by minimising Eq. (\ref{eq:min_function}) to the one found  by deprojecting the lensing potential. However as the gravitational potential is not uniquely defined it is more favourable to compare the cumulative masses defined in Eqs. (\ref{eq:cum_mass_gas}) and ({\ref{eq:cum_mass_lensing}). If the cluster is exactly in hydrostatic equilibrium, so that Eq. (\ref{eq:equilibrium_condition}) is satisfied, the masses should be identical for all distances $r$ from the cluster centre except for small deviations caused by reconstruction errors. Otherwise differences between the masses directly reflect the differences between the gravitational field and $\vec{\nabla} p/\rho$. In the next two sections we test this method to probe hydrostatic equilibrium in galaxy clusters by performing such a comparison using synthetic observation of analytically modelled and numerically simulated clusters.

\section{Probing hydrostatic equilibrium in analytically modelled clusters}
\label{sec:analytic}

\subsection{The analytic halo model}

We use an analytic halo model with a NFW total (gas+DM) density profile to test the methods introduced in section \ref{sec:techniques}. Thus the total matter density $\rho_m$ and the gravitational potential are given by
\begin{align}
 &\rho_m = \frac{c}{\frac{r}{r_s}(1+\frac{r}{r_s})^2}, \\
 &\phi = \frac{4 \pi G c r_s^3}{r} \ln\big(\frac{r_s}{r+r_s}\big), \label{eq:analytic_pot}
\end{align}
where $r_s$ is the NFW scaling radius and $c = 4\rho(r_s)$ fixes the normalisation of the density profile. For the cluster gas we assume in this toy model that the ratio $f$ of $\vec{\nabla} p/\rho$ to $-\vec{\nabla{\phi}}$ is constant but can be different from 1. So Eq. (\ref{eq:equilibrium_condition}) generalises to
\begin{equation}
    f \, \vec{\nabla} \phi = -\frac{\vec{\nabla} p}{\rho}.
\end{equation}
We further assume a polytropic equation of state $T \propto \rho^{\gamma-1}$ for the cluster gas, where $T$ is the gas temperature and $\gamma$ the polytropic index. Then the gas density $\rho$ and temperature $T$ satisfy
\begin{align}
 &\rho \propto \left[\frac{(1-\gamma)\phi}{\gamma}\right]^{\frac{1}{\gamma-1}}, \\
 &kT = f \frac{(1-\gamma)\phi}{\gamma}\bar{m},
\end{align}
where $k$ is Boltzmann's constant and $\bar{m}$ is the mean gas particle mass. In the following, we adopt $\gamma=1.2$, which is consistent with X-ray temperature profiles of nearby clusters \citep[]{1998ApJ...503...77M}, and fix the normalisation of $\rho$ by requiring a baryon fraction of 0.12 at the scale radius, which we set to $r_s=300h^{-1}\mathrm{kpc}$. Note that the lengths here and below are given in comoving units. A reduced Hubble parameter of $h=0.7$ is used and $c$ is chosen to be $1.9\times10^6 h^{-1} M_{\odot}/(h^{-1} \mathrm{kpc})^3$. These choices for $r_s$ and $c$ correspond to a massive galaxy cluster. To test the reconstruction methods we put this analytically modelled cluster at a redshift of $z=0.3$ and produce synthetic X-ray, thermal SZ and lensing observations. 

\subsection{Synthetic observations}
\label{sec:observations}

For the X-ray observations we use the table of the cooling function discussed in section \ref{sec:gas_reconstructions} and a $128 \times 128 \times 128$ grid with $1.5 h^{-1} \mathrm{Mpc}$ side length to project the gas distribution and get a map of the X-ray surface brightness in a 0.25-7.0 keV band. Then, except for reconstructions we specifically characterise as done without observational noise, we add photon noise corresponding to $10^4$ observed source photons to these maps using the same method as in Paper I. 

The thermal SZ maps are generated like in Paper I by projecting the product of cluster gas density and temperature along the line-of-sight onto a $128 \times 128$ grid with $1.5 h^{-1} \mathrm{Mpc}$ side length. Then the result is appropriately rescaled and, unless stated otherwise, noise corresponding to future ALMA Band 3 observations is added. In Band 3 (84-116 GHz) and in its compact configuration, ALMA will be able to achieve a temperature sensitivity of $50 \mu\textrm{K}$ at a spatial resolution of $\sim 3$ arcsec in about four hours of observation \citep[see][]{BU99.1}. 

To produce maps of the lensing potential we project the mass inside a cube of $6 \, h^{-1} \mathrm{Mpc}$ side length which is centred on the cluster along the line-of-sight and calculate the convergence. The grid we use for this purpose is chosen such that each pixel corresponds to roughly $1/3$ square arcminute on the sky, so that it contains about 10 galaxies, if we assume an average density of background galaxies useable for a weak lensing analysis of $n_g=30/\mathrm{arcmin}^2$. For instance for a cluster at redshift $z=0.3$ a $44 \times 44$ convergence map covers the projection of the  $6 \, h^{-1} \mathrm{Mpc}$ cube on the sky. 

For lensing reconstructions with observational noise, normally distributed noise with variance
\begin{equation}
  \sigma_\kappa^2 = \frac{\sigma^2 \sigma_\epsilon^2}{\pi n_g a^4}
      \Big(1-\exp\big(-\frac{a^2}{2\sigma^2}\big)-\sqrt{\frac{\pi}{2}} \frac{a}{\sigma}\,\mathrm{erf}\big(\frac{a}{\sqrt{2}\sigma}\big)\Big)^2,                  
\end{equation}
is added to each pixel of the convergence map. Here $\sigma_\kappa^2$ is the variance expected for a weak lensing reconstruction of the convergence for a density $n_g$ of background galaxies with an intrinsic ellipticity dispersion $\sigma_\epsilon$, and for an angular pixel size $a$ of the convergence map \citep[see][]{VW00.1}. It is assumed that the galaxy ellipticities are smoothed with a Gaussian of angular standard deviation $\sigma$ before the reconstruction. We choose $\sigma = a$ and $\sigma_\epsilon = 0.3$.

Then the convergence map is used to calculate the lensing potential in Fourier space. A source redshift of 1.5 is assumed. To reduce errors introduced by the implicit assumption of a periodic convergence field of such Fourier methods \citep[see e.g.][]{PU05.1}, we first zero-pad the convergence map to $2048 \times 2048$ pixels. Once the map of the lensing potential is calculated, we crop it to its original size before using it for reconstructions.

\subsection{Cumulative mass profiles}

We applied these methods to produce synthetic X-ray, SZ and lensing observation for the analytic halo described above. We then generate three-dimensional reconstructions of the cluster gas based on X-ray and SZ data and reconstructions of the gravitational potential based on lensing data using the methods detailed in section \ref{sec:techniques} and in Paper I. 

Figure \ref{fig:analytic} shows the cumulative mass profiles $M_{<r,\,\mathrm{XSZ}}(r)$, obtained from the XSZ reconstructions, and $M_{<r,\,\mathrm{lensing}}(r)$, obtained from the lensing reconstructions, and compares them to the original analytic profile. The profiles are shown for reconstructions based on data without observational noise and for reconstructions based on noisy data and for inclination angles $i=30^\circ$ and $i=70^\circ$ between the symmetry axis and the line-of-sight. The inclination angles were assumed to be known for the reconstructions. See Paper I for methods to determine them from the observations. For the gas reconstructions ratios $f$ of $\vec{\nabla} p/\rho$ to $-\vec{\nabla{\phi}}$ of $f=1.0$ and $f=0.8$ were used. The lensing reconstructions were done using both flat priors and spherically symmetrised priors.

The XSZ and the lensing mass profiles agree very well for a halo in hydrostatic equilibrium ($f=1.0$) and when using data without noise and an inclination $i=70^\circ$ (see upper right panel of Fig. \ref{fig:analytic}). They also excellently match the original analytic profile. The only significant difference between the profiles is that the lensing mass is too small very close to the cluster centre. However, this is completely expected because the lensing observations lack the resolution required to accurately resolve this region. When perturbing the hydrostatic equilibrium by 20\%, in other words when assuming $f=0.8$, the XSZ reconstructed mass profile is essentially 0.8 times the original analytic profile as theoretically expected. For such a halo one can easily see a significant ($\sim 20\%$) difference between the lensing and XSZ mass profiles, which directly reflects the deviation from hydrostatic equilibrium. Also when adding noise to the synthetic observations (see lower right panel) such a deviation from hydrostatic equilibrium can be faithfully reproduced. For smaller inclination angles of $i=30^\circ$ (left panels) the accuracy of the reconstructions is somewhat lower and one can also see significant differences between the lensing reconstructions based on a flat prior and a spherically symmetrised prior. The latter reproduce the original profiles much better.  Thus, for such small inclinations one can detect deviations from hydrostatic equilibrium by comparing lensing reconstructions based on a spherically symmetrised prior to XSZ reconstructions. Also note that for a randomly oriented cluster sample only about 13\% of the clusters have inclination angles smaller than $30^\circ$. Paper I contains a more detailed discussion of the dependence of the reconstruction accuracy on the inclination angle.

\begin{figure*}
{
\begin{picture}(0,0)%
\scalebox{1.4}{\includegraphics{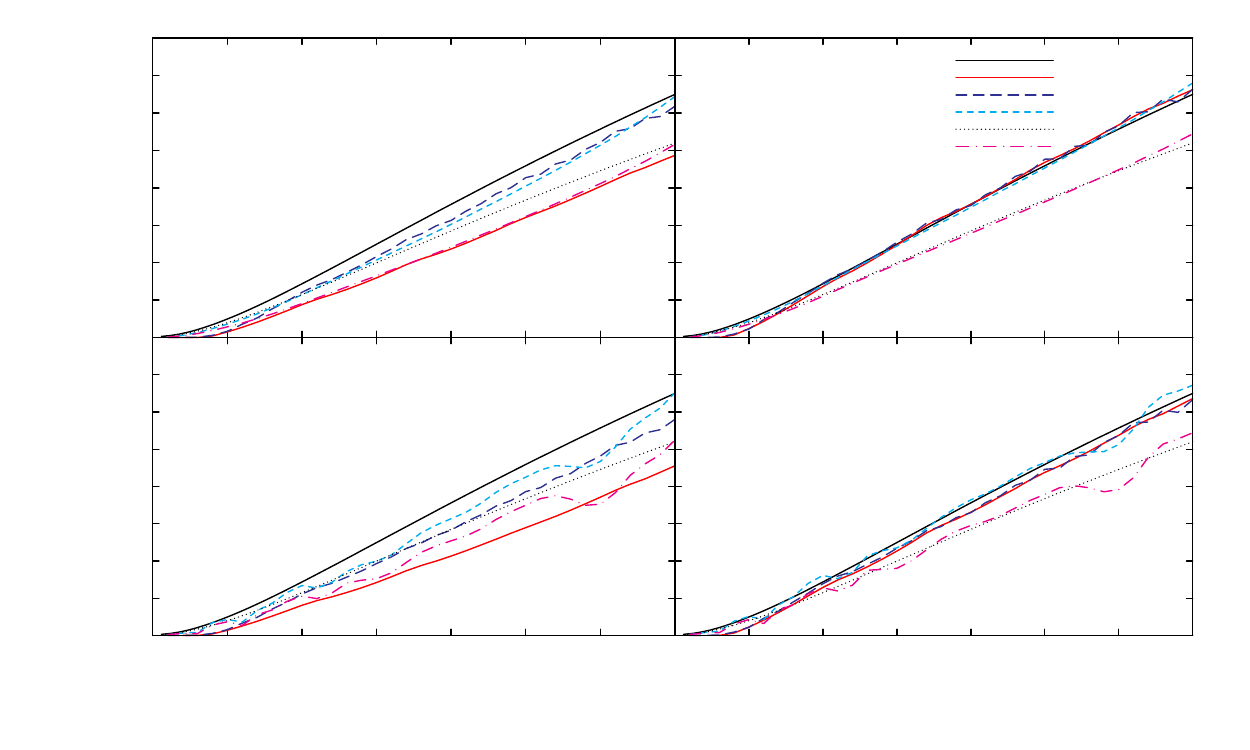}}%
\end{picture}%
\begingroup
\setlength{\unitlength}{0.0280bp}%
\begin{picture}(18000,10800)(0,0)%
\put(9445,5940){\makebox(0,0)[r]{\strut{}}}%
\put(9445,6479){\makebox(0,0)[r]{\strut{}}}%
\put(9445,7017){\makebox(0,0)[r]{\strut{}}}%
\put(9445,7556){\makebox(0,0)[r]{\strut{}}}%
\put(9445,8095){\makebox(0,0)[r]{\strut{}}}%
\put(9445,8634){\makebox(0,0)[r]{\strut{}}}%
\put(9445,9172){\makebox(0,0)[r]{\strut{}}}%
\put(9445,9711){\makebox(0,0)[r]{\strut{}}}%
\put(9445,10250){\makebox(0,0)[r]{\strut{}}}%
\put(9720,5390){\makebox(0,0){\strut{}}}%
\put(10785,5390){\makebox(0,0){\strut{}}}%
\put(11850,5390){\makebox(0,0){\strut{}}}%
\put(12915,5390){\makebox(0,0){\strut{}}}%
\put(13980,5390){\makebox(0,0){\strut{}}}%
\put(15045,5390){\makebox(0,0){\strut{}}}%
\put(16110,5390){\makebox(0,0){\strut{}}}%
\put(14513,6479){\makebox(0,0)[l]{\strut{}$i=70^\circ$, no noise}}%
\put(13492,9927){\makebox(0,0)[r]{\strut{} \tiny original analytic mass}}%
\put(13492,9680){\makebox(0,0)[r]{\strut{} \tiny lensing reconstr., flat prior}}%
\put(13492,9433){\makebox(0,0)[r]{\strut{} \tiny lensing reconstr., spher. prior}}%
\put(13492,9186){\makebox(0,0)[r]{\strut{} \tiny f=1.0 XSZ reconstr.}}%
\put(13492,8939){\makebox(0,0)[r]{\strut{} \tiny 0.8 x original analytic mass}}%
\put(13492,8692){\makebox(0,0)[r]{\strut{} \tiny f=0.8 XSZ reconstr.}}%
\put(1925,9711){\makebox(0,0)[r]{\strut{}3.5}}%
\put(1925,9172){\makebox(0,0)[r]{\strut{}3}}%
\put(1925,8634){\makebox(0,0)[r]{\strut{}2.5}}%
\put(1925,8095){\makebox(0,0)[r]{\strut{}2}}%
\put(1925,7556){\makebox(0,0)[r]{\strut{}1.5}}%
\put(1925,7017){\makebox(0,0)[r]{\strut{}1}}%
\put(1925,6479){\makebox(0,0)[r]{\strut{}0.5}}%
\put(1925,5940){\makebox(0,0)[r]{\strut{}0}}%
\put(2200,5390){\makebox(0,0){\strut{}}}%
\put(3274,5390){\makebox(0,0){\strut{}}}%
\put(4349,5390){\makebox(0,0){\strut{}}}%
\put(5423,5390){\makebox(0,0){\strut{}}}%
\put(6497,5390){\makebox(0,0){\strut{}}}%
\put(7571,5390){\makebox(0,0){\strut{}}}%
\put(8646,5390){\makebox(0,0){\strut{}}}%
\put(550,8095){\rotatebox{90}{\makebox(0,0){\strut{} $M_{<r} [10^{14} h^{-1} M_{\odot}]$}}}%
\put(7034,6479){\makebox(0,0)[l]{\strut{}$i=30^\circ$, no noise}}%
\put(9445,1650){\makebox(0,0)[r]{\strut{}}}%
\put(9445,2186){\makebox(0,0)[r]{\strut{}}}%
\put(9445,2722){\makebox(0,0)[r]{\strut{}}}%
\put(9445,3259){\makebox(0,0)[r]{\strut{}}}%
\put(9445,3795){\makebox(0,0)[r]{\strut{}}}%
\put(9445,4331){\makebox(0,0)[r]{\strut{}}}%
\put(9445,4867){\makebox(0,0)[r]{\strut{}}}%
\put(9445,5404){\makebox(0,0)[r]{\strut{}}}%
\put(9445,5940){\makebox(0,0)[r]{\strut{}}}%
\put(9720,1100){\makebox(0,0){\strut{} 0}}%
\put(10785,1100){\makebox(0,0){\strut{} 100}}%
\put(11850,1100){\makebox(0,0){\strut{} 200}}%
\put(12915,1100){\makebox(0,0){\strut{} 300}}%
\put(13980,1100){\makebox(0,0){\strut{} 400}}%
\put(15045,1100){\makebox(0,0){\strut{} 500}}%
\put(16110,1100){\makebox(0,0){\strut{} 600}}%
\put(13447,375){\makebox(0,0){\strut{}distance $r$ from cluster centre [$h^{-1}$ kpc]}}%
\put(14513,2186){\makebox(0,0)[l]{\strut{}$i=70^\circ$, noise}}%
\put(1925,5404){\makebox(0,0)[r]{\strut{}3.5}}%
\put(1925,4867){\makebox(0,0)[r]{\strut{}3}}%
\put(1925,4331){\makebox(0,0)[r]{\strut{}2.5}}%
\put(1925,3795){\makebox(0,0)[r]{\strut{}2}}%
\put(1925,3259){\makebox(0,0)[r]{\strut{}1.5}}%
\put(1925,2722){\makebox(0,0)[r]{\strut{}1}}%
\put(1925,2186){\makebox(0,0)[r]{\strut{}0.5}}%
\put(1925,1650){\makebox(0,0)[r]{\strut{}0}}%
\put(2200,1100){\makebox(0,0){\strut{} 0}}%
\put(3274,1100){\makebox(0,0){\strut{} 100}}%
\put(4349,1100){\makebox(0,0){\strut{} 200}}%
\put(5423,1100){\makebox(0,0){\strut{} 300}}%
\put(6497,1100){\makebox(0,0){\strut{} 400}}%
\put(7571,1100){\makebox(0,0){\strut{} 500}}%
\put(8646,1100){\makebox(0,0){\strut{} 600}}%
\put(550,3795){\rotatebox{90}{\makebox(0,0){\strut{} $M_{<r} [10^{14} h^{-1} M_{\odot}$]}}}%
\put(5960,375){\makebox(0,0){\strut{}distance $r$ from cluster centre [$h^{-1}$ kpc]}}%
\put(7034,2186){\makebox(0,0)[l]{\strut{}$i=30^\circ$, noise}}%
\end{picture}%
\endgroup
}
\caption{Cumulative mass profiles $M_{<r}(r)$ of an analytic halo and its reconstructions from X-ray and SZ maps with and without observational noise as well as from lensing maps with and without noise. The upper panels show the results obtained from maps without noise, while the lower panels show the profiles found from noisy maps. For the reconstructions shown in the left panels an inclination angle of $30^\circ$ was used and assumed to be known, while the right panels show the corresponding results for an inclination angle of $70^\circ$. Lensing reconstructions are shown for a flat and for the spherically symmetrised prior. The XSZ reconstructions were done for halos with ratios $f$ of $\vec{\nabla} p/\rho$ to $-\vec{\nabla{\phi}}$ of $1.0$ and 0.8. For comparison we also show the original analytic cumulative mass multiplied by 0.8.}
\label{fig:analytic}
\end{figure*}
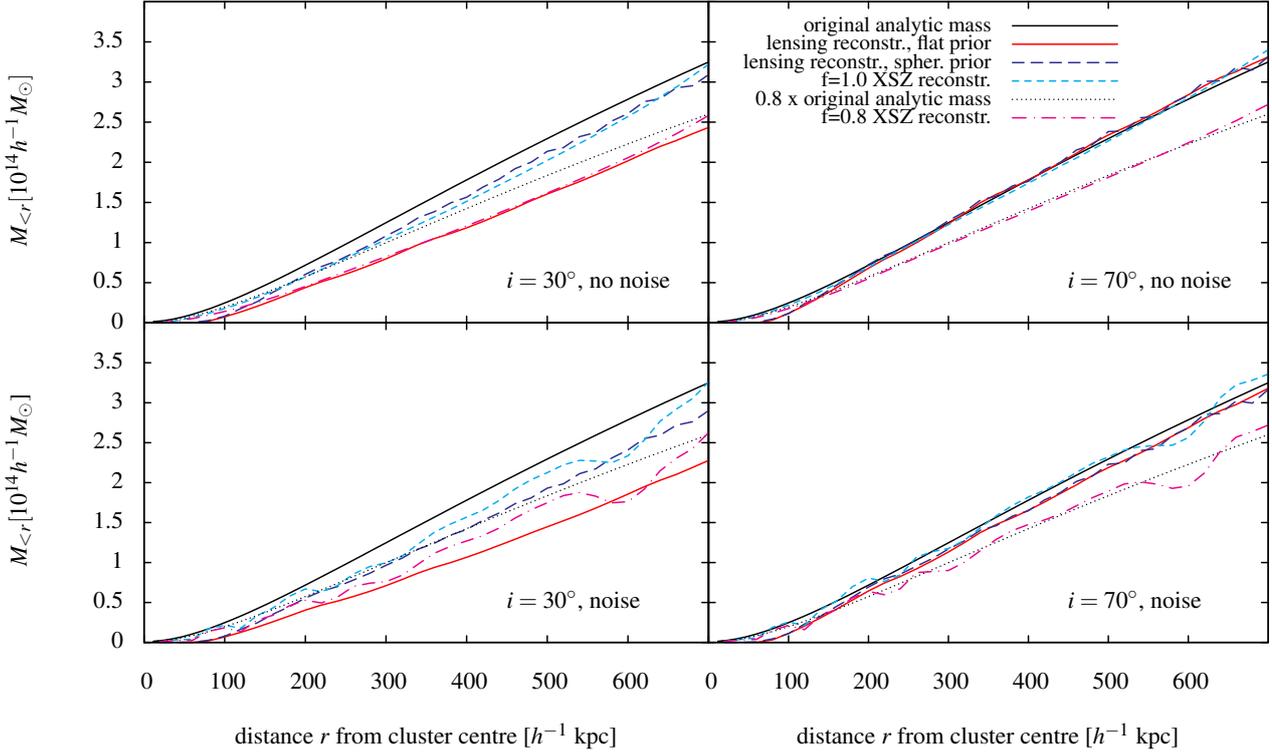

Above we used spherically symmetrised priors in the lensing reconstruction of spherically symmetric halos. It is reassuring, but not really surprising that this works well. We thus need to check whether or not a spherically symmetrised prior also improves the lensing reconstruction quality of elliptical halos for small inclination angles. In Figure \ref{fig:analytic_ellip} we show lensing reconstructions of the cumulative mass profile of an elliptic analytic halo with an NFW density profile but isodensity surfaces that are prolate spheroids with a major to minor axis ratio of 2 to 1. The lensing reconstructions with a spherically symmetrised prior reproduce the original analytic profile well, both for small and for large inclination angles. On the other hand when using a flat prior we again obtain too small lensing masses for small inclination angles. It is thus favourable to use a spherically symmetrised prior for the iterative deprojection of the lensing potential. 
 
\begin{figure}
\scalebox{0.72}
{
\begin{picture}(0,0)%
\includegraphics{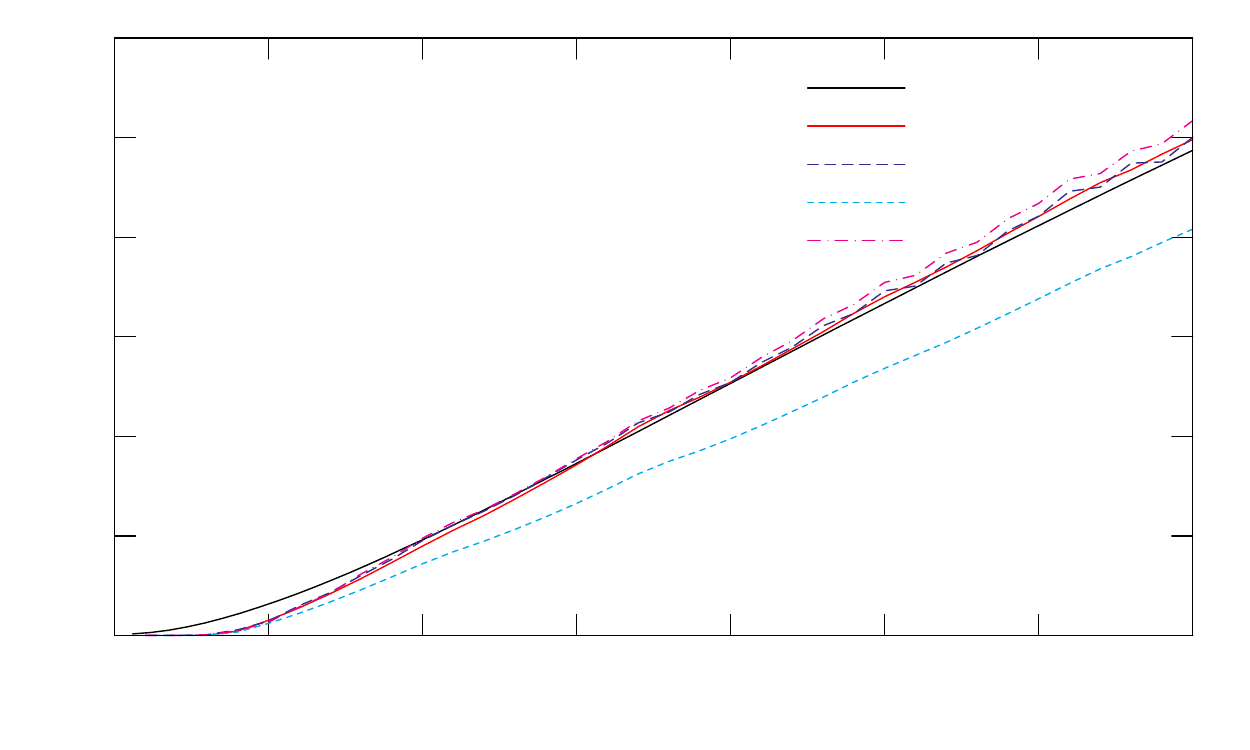}%
\end{picture}%
\begingroup
\setlength{\unitlength}{0.0200bp}%
\begin{picture}(18000,10800)(0,0)%
\put(1375,10250){\makebox(0,0)[r]{\strut{}6}}%
\put(1375,8817){\makebox(0,0)[r]{\strut{}5}}%
\put(1375,7383){\makebox(0,0)[r]{\strut{}4}}%
\put(1375,5950){\makebox(0,0)[r]{\strut{}3}}%
\put(1375,4517){\makebox(0,0)[r]{\strut{}2}}%
\put(1375,3083){\makebox(0,0)[r]{\strut{}1}}%
\put(1375,1650){\makebox(0,0)[r]{\strut{}0}}%
\put(1650,1100){\makebox(0,0){\strut{} 0}}%
\put(3868,1100){\makebox(0,0){\strut{} 100}}%
\put(6086,1100){\makebox(0,0){\strut{} 200}}%
\put(8304,1100){\makebox(0,0){\strut{} 300}}%
\put(10521,1100){\makebox(0,0){\strut{} 400}}%
\put(12739,1100){\makebox(0,0){\strut{} 500}}%
\put(14957,1100){\makebox(0,0){\strut{} 600}}%
\put(17175,1100){\makebox(0,0){\strut{} 700}}%
\put(550,5950){\rotatebox{90}{\makebox(0,0){\strut{}$M_{<r} [10^{14} h^{-1} M_{\odot}]$}}}%
\put(9412,275){\makebox(0,0){\strut{}distance $r$ from cluster centre [$h^{-1}$ kpc]}}%
\put(11355,9533){\makebox(0,0)[r]{\strut{}original analytic mass}}%
\put(11355,8983){\makebox(0,0)[r]{\strut{}$i=70^\circ$, lensing reconstr., flat prior}}%
\put(11355,8433){\makebox(0,0)[r]{\strut{}$i=70^\circ$, lensing reconstr., spher. prior}}%
\put(11355,7883){\makebox(0,0)[r]{\strut{}$i=30^\circ$, lensing reconstr., flat prior}}%
\put(11355,7333){\makebox(0,0)[r]{\strut{}$i=30^\circ$, lensing reconstr., spher. prior}}%
\end{picture}%
\endgroup
}
\caption{Cumulative mass profiles $M_{<r}(r)$ of an ellipsoidal analytic halo and its reconstructions from lensing maps without noise. Inclination angles of $i=30^\circ$ and $i=70^\circ$ were used for the synthetic observations and assumed to be known for the reconstructions. Lensing reconstructions are shown for a flat and for the spherically symmetrised prior.}
\label{fig:analytic_ellip}
\end{figure}
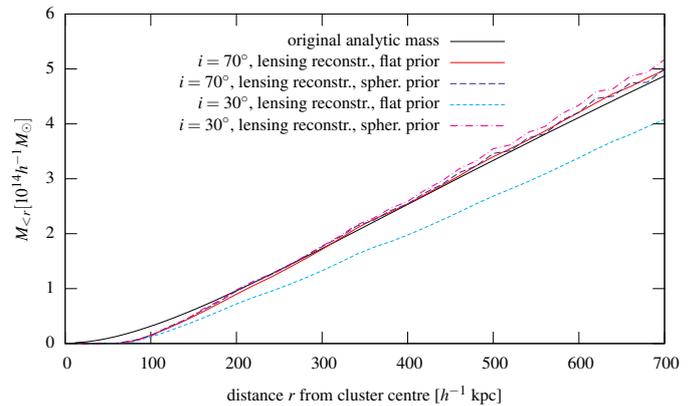

\subsection{Gravitational potential reconstructions}

In Figure \ref{fig:analytic_pots} we show three-dimensional reconstructions of the gravitational potential of the analytic halo from X-ray and SZ data and from lensing data as well as the original analytic gravitational potential described by Eq. (\ref{eq:analytic_pot}). Reconstructions that are based on idealised observations without noise and on more realistic noisy observations are shown. The XSZ potential reconstructions were obtained from the X-ray, SZ cluster gas reconstructions by assuming hydrostatic equilibrium and by using the minimisation method described in section \ref{sec:x-ray_sz_pot_reconstruction}. The lensing reconstructions were obtained directly by deprojecting the lensing potential. The XSZ reconstructions reproduce the inner region of the cluster well, while the lensing reconstructions lack the resolution to accurately resolve this innermost part. Between distances $r$ from the cluster centre of $150  h^{-1}\mathrm{kpc}$ and $450 h^{-1}\mathrm{kpc}$ both reconstruction methods yield very good results. Farther outside the lensing reconstruction is still accurate, while the XSZ reconstruction becomes more and more unrealistic. This is partly due to different noise properties. But in the example shown in Figure \ref{fig:analytic_pots} it is also due to the smaller box size used for the XSZ reconstructions and reconstruction artefacts that develop close to the perimeter of this box. The weighting function $w(r)$, which was introduced to prevent non-local effects of these artefacts, was chosen  to decrease from unity to zero between $r=450 h^{-1}\mathrm{kpc}$ and $r=600 h^{-1}\mathrm{kpc}$ in these XSZ reconstructions. Thus they become unrealistic farther outside. 

\begin{figure*}
\scalebox{0.65}
{
\begin{picture}(0,0)%
\ifpdf
	\includegraphics{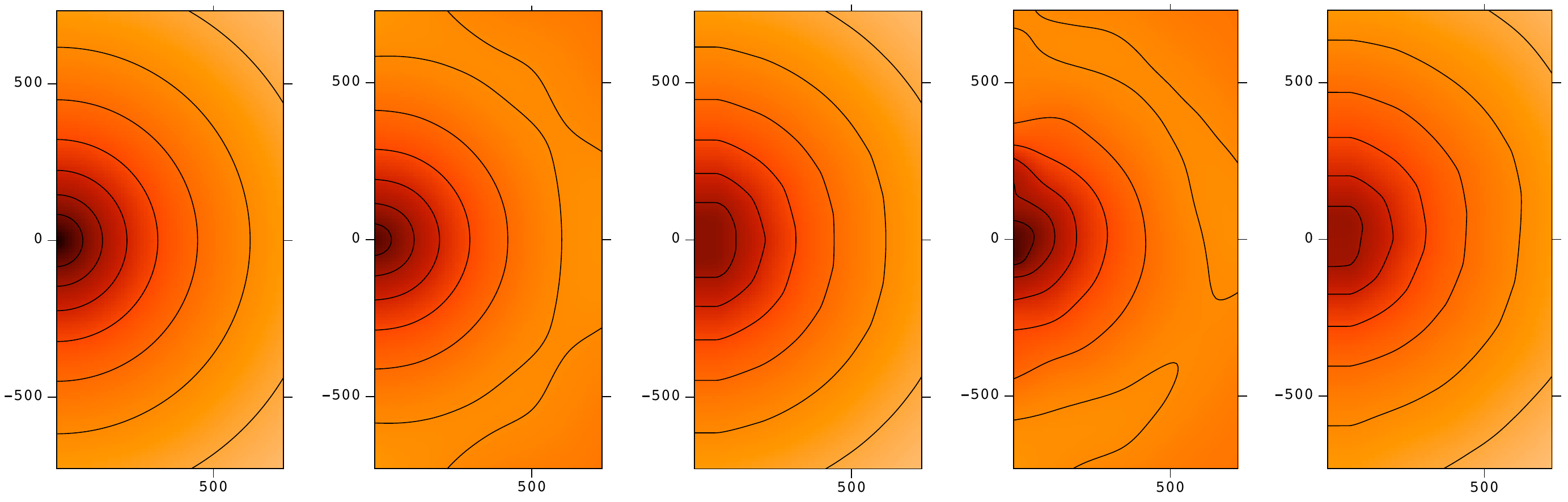}
\else
	\includegraphics[bb =  25 263 350 519, clip=true]{pots_no_noise.ps}%
	\includegraphics[bb =  400 263 575 519, clip=true]{pots_no_noise.ps}%
	\includegraphics[bb =  200 263 350 519, clip=true]{pots_noise.ps}%
	\includegraphics[bb =  400 263 575 519, clip=true]{pots_noise.ps}%
\fi
\end{picture}%
\setlength{\unitlength}{0.0200bp}%
\begingroup
\begin{picture}(18000,13200)(0,0)
\ifpdf
	\put(350,6500){\rotatebox{90}{\makebox(0,0){\large \strut{$Z [h^{-1}$ kpc]}}}}
	\put(8350,6500){\rotatebox{90}{\makebox(0,0){\large \strut{$Z [h^{-1}$ kpc]}}}}	
	\put(16350,6500){\rotatebox{90}{\makebox(0,0){\large \strut{$Z [h^{-1}$ kpc]}}}}
	\put(24350,6500){\rotatebox{90}{\makebox(0,0){\large \strut{$Z [h^{-1}$ kpc]}}}}
	\put(32350,6500){\rotatebox{90}{\makebox(0,0){\large \strut{$Z [h^{-1}$ kpc]}}}}
	\put(3500,250){\makebox(0,0){\large \strut{$R [h^{-1}$ kpc]}}}
	\put(11500,250){\makebox(0,0){\large \strut{$R [h^{-1}$ kpc]}}}
	\put(19500,250){\makebox(0,0){\large \strut{$R [h^{-1}$ kpc]}}}
	\put(27500,250){\makebox(0,0){\large \strut{$R [h^{-1}$ kpc]}}}
	\put(35500,250){\makebox(0,0){\large \strut{$R [h^{-1}$ kpc]}}}
	\put(4000,12950){\makebox(0,0){\large \strut{analytic}}}
	\put(12000,12950){\makebox(0,0){\large \strut{XSZ reconstr., no noise}}}
	\put(20000,12950){\makebox(0,0){\large \strut{lensing reconstr., no  noise}}}
	\put(28000,12950){\makebox(0,0){\large \strut{XSZ reconstr., noise}}}
	\put(36000,12950){\makebox(0,0){\large \strut{lensing reconstr., noise}}}
\else
    \put(350,6700){\rotatebox{90}{\makebox(0,0){\large \strut{$Z [h^{-1}$ kpc]}}}}
	\put(8450,6700){\rotatebox{90}{\makebox(0,0){\large \strut{$Z [h^{-1}$ kpc]}}}}	
	\put(16550,6700){\rotatebox{90}{\makebox(0,0){\large \strut{$Z [h^{-1}$ kpc]}}}}
	\put(24650,6700){\rotatebox{90}{\makebox(0,0){\large \strut{$Z [h^{-1}$ kpc]}}}}
	\put(32750,6700){\rotatebox{90}{\makebox(0,0){\large \strut{$Z [h^{-1}$ kpc]}}}}
	\put(3400,350){\makebox(0,0){\large \strut{$R [h^{-1}$ kpc]}}}
	\put(11500,350){\makebox(0,0){\large \strut{$R [h^{-1}$ kpc]}}}
	\put(19600,350){\makebox(0,0){\large \strut{$R [h^{-1}$ kpc]}}}
	\put(27700,350){\makebox(0,0){\large \strut{$R [h^{-1}$ kpc]}}}
	\put(35800,350){\makebox(0,0){\large \strut{$R [h^{-1}$ kpc]}}}
	\put(4200,13150){\makebox(0,0){\large \strut{analytic}}}
	\put(12400,13150){\makebox(0,0){\large \strut{XSZ reconstr., no noise}}}
	\put(20500,13150){\makebox(0,0){\large \strut{lensing reconstr., no  noise}}}
	\put(28700,13150){\makebox(0,0){\large \strut{XSZ reconstr., noise}}}
	\put(36800,13150){\makebox(0,0){\large \strut{lensing reconstr., noise}}}
\fi
\end{picture}
\endgroup
}
\caption{Gravitational potential of an analytic halo and its reconstructions from synthetic X-ray and SZ observations and from synthetic lensing observations each with and without observational noise. The XSZ reconstructions work well even close to the cluster centre, where lensing observations lack the resolution to accurately resolve the central peak. However, farther outside the lensing reconstructions perform better, due to the different noise properties, but in the example shown here also because of the smaller box size used for the XSZ reconstructions.}
\label{fig:analytic_pots}
\end{figure*}

\section{Probing the dynamical state of numerically simulated clusters} 
\label{sec:numerical}

\subsection{The simulations}

In the previous section, novel methods to probe hydrostatic equilibrium in clusters of galaxies were tested with synthetic observations of analytically modelled clusters. For a more realistic test we now apply these methods to a sample of four numerically simulated galaxy clusters. The same sample was also used in Paper I. The simulations were carried out by Klaus Dolag with the GADGET-2 code \citep[]{SP05.1}, a new version of the parallel TreeSPH simulation code GADGET \citep[]{SP01.1}. The cluster regions were extracted from a dissipation-less (dark matter only) simulation with a box size of $479h^{-1}$ Mpc of a flat $\Lambda$CDM model with $\Omega_m=0.3$, $h=0.7$, $\sigma_8=0.9$ \citep[see][]{YO01.1}. They were re-simulated with higher resolution using the ``Zoomed Initial Conditions'' (ZIC) technique \citep[]{TO97.1}. Gas was introduced into the high-resolution region by splitting each parent particle into a gas and a dark matter particle, which were then displaced by half the mean inter-particle distance, such that the centre-of-mass and the momentum were conserved. The mass ratio of gas to dark matter particles was set to obtain $\Omega_b=0.04$. The final mass resolution was $m_{\rm DM}=1.13\times 10^9\:h^{-1}M_\odot$ and $m_{\rm gas}=1.7\times 10^8\:h^{-1}M_\odot$ for dark-matter and gas particles within the high-resolution region, respectively. The simulations we use follow the dynamics of the dark matter and the adiabatic evolution of the cluster gas, but they ignore radiative cooling. They are described in more detail in \cite{PU05.1} and \cite{DO05.1}.

\subsection{Cumulative mass profiles and hydrostatic equilibrium in simulated clusters}

We produced synthetic X-ray, thermal SZ and lensing observations of these four simulated clusters for 28 simulation snapshots between redshifts 0.58 and 0.1 and three lines-of-sight using essentially the same methods as in section \ref{sec:observations} for the analytic halo. The only difference is that we did not use a three-dimensional grid for projections along the line-of-sight. For the X-ray and SZ maps the X-ray luminosities and integrated Compton $y$ parameters of the gas particles are projected directly onto a two-dimensional $128\times128$ grid by using the particles' projected SPH smoothing kernel. The convergence of the simulated clusters is found in a similar way by projecting the masses of both gas and dark matter particles onto a two-dimensional grid, whose dimensions are again chosen such that one pixel corresponds to roughly $1/3$ square arcminute on the sky. Observational noise is added in exactly the same way as in section \ref{sec:observations}.

Based on these synthetic observations we perform three-dimensional XSZ reconstructions of the cluster gas distribution and lensing reconstructions of the gravitational potential. The inclination angle is assumed to be known for the reconstructions. Paper I discusses how it can be determined from data and how a symmetry axis is chosen for the simulated clusters. Spherically symmetrised priors are used for the lensing reconstructions. The reconstructions are then used to probe hydrostatic equilibrium by calculating and comparing their cumulative mass profiles $M_{<r,\,\mathrm{XSZ}}(r)$ and $M_{<r,\,\mathrm{lensing}}(r)$. 

In Figure \ref{fig:simulated_relaxed} we show these profiles for two clusters that did not experience a major merger recently. For comparison we also show the original simulated mass profile and the profile that would be expected from the original simulated gas distribution by assuming hydrostatic equilibrium. The latter is calculated like $M_{<r,\,\mathrm{XSZ}}(r)$, however directly from the simulated gas distribution rather than the reconstructed one. We again use 128 rings which are equally spaced in the polar angle $\theta$ to numerically evaluate the surface integral in Eq. (\ref{eq:cum_mass_gas}), but as the simulated gas distribution is not perfectly axisymmetric we use 128 sampling points equally spaced in the longitude angle for each of these rings. The gas density $\rho$ and the pressure gradient $\vec{\nabla} p$ are calculated at each sampling point using the SPH formalism, i.e. by summing up the contributions from all nearby particles using their SPH smoothing kernels and the gradients thereof. For these relaxed clusters the XSZ reconstructed profiles and the lensing reconstructed profiles agree well with each other, with the original mass profile and the profile obtained from the original gas distribution. This shows that for such relaxed clusters this method allows accurate and consistent lensing and XSZ mass estimates. The results also confirm that these clusters are close to hydrostatic equilibrium. 

\begin{figure*}
{
\begin{picture}(0,0)%
\scalebox{1.4}{\includegraphics{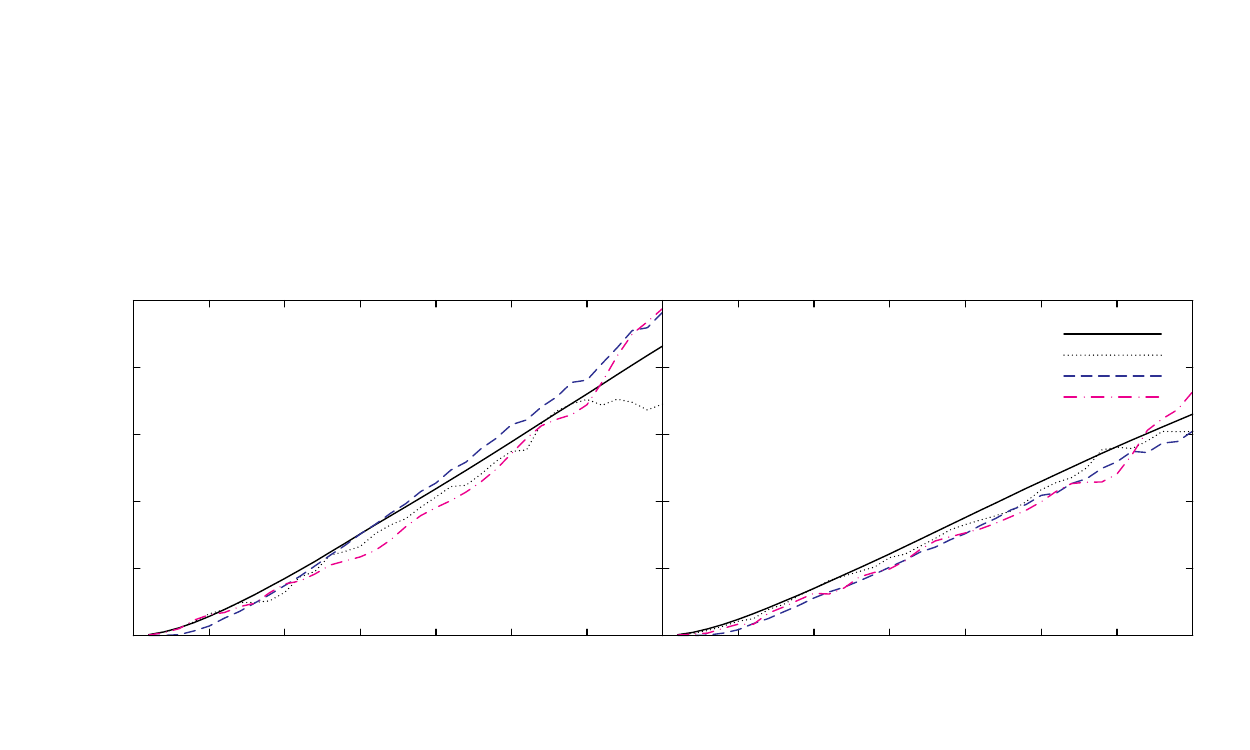}}%
\end{picture}%
\begingroup
\setlength{\unitlength}{0.0280bp}%
\begin{picture}(18000,6800)(0,0)%
\put(1650,1650){\makebox(0,0)[r]{\strut{} 0}}%
\put(1650,2614){\makebox(0,0)[r]{\strut{} 1}}%
\put(1650,3578){\makebox(0,0)[r]{\strut{} 2}}%
\put(1650,4542){\makebox(0,0)[r]{\strut{} 3}}%
\put(1650,5506){\makebox(0,0)[r]{\strut{} 4}}%
\put(1650,6470){\makebox(0,0)[r]{\strut{} 5}}%
\put(1925,1100){\makebox(0,0){\strut{} 0}}%
\put(3013,1100){\makebox(0,0){\strut{} 100}}%
\put(4101,1100){\makebox(0,0){\strut{} 200}}%
\put(5189,1100){\makebox(0,0){\strut{} 300}}%
\put(6276,1100){\makebox(0,0){\strut{} 400}}%
\put(7364,1100){\makebox(0,0){\strut{} 500}}%
\put(8452,1100){\makebox(0,0){\strut{} 600}}%
\put(550,4060){\rotatebox{90}{\makebox(0,0){\strut{} $M_{<r} [10^{14} h^{-1} M_{\odot}]$}}}%
\put(5732,375){\makebox(0,0){\strut{}distance r from cluster centre $[h^{-1}$ kpc]}}%
\put(6603,2421){\makebox(0,0)[l]{\strut{} cluster g1 z=0.25}}%
\put(9265,1650){\makebox(0,0)[r]{\strut{}}}%
\put(9265,2614){\makebox(0,0)[r]{\strut{}}}%
\put(9265,3578){\makebox(0,0)[r]{\strut{}}}%
\put(9265,4542){\makebox(0,0)[r]{\strut{}}}%
\put(9265,5506){\makebox(0,0)[r]{\strut{}}}%
\put(9265,6470){\makebox(0,0)[r]{\strut{}}}%
\put(9540,1100){\makebox(0,0){\strut{} 0}}%
\put(10631,1100){\makebox(0,0){\strut{} 100}}%
\put(11721,1100){\makebox(0,0){\strut{} 200}}%
\put(12812,1100){\makebox(0,0){\strut{} 300}}%
\put(13903,1100){\makebox(0,0){\strut{} 400}}%
\put(14994,1100){\makebox(0,0){\strut{} 500}}%
\put(16084,1100){\makebox(0,0){\strut{} 600}}%
\put(13357,375){\makebox(0,0){\strut{}distance r from cluster centre $[h^{-1}$ kpc]}}%
\put(14230,2421){\makebox(0,0)[l]{\strut{}cluster g51 z=0.3}}%
\put(15046,5988){\makebox(0,0)[r]{\strut{}original simulated mass}}%
\put(15046,5686){\makebox(0,0)[r]{\strut{}from simulated gas distribution}}%
\put(15046,5384){\makebox(0,0)[r]{\strut{}lensing reconstr.}}%
\put(15046,5082){\makebox(0,0)[r]{\strut{}XSZ reconstr.}}%
\end{picture}%
\endgroup
}
\caption{Cumulative mass profiles $M_{<r}(r)$ of relaxed simulated clusters g1 at redshift  $z=0.25$ and g51 at redshift $z=0.3$. Profiles of  the original simulated mass distribution, of the lensing and of the XSZ reconstructions are shown, as well as the profile obtained directly from the simulated gas distribution by assuming hydrostatic equilibrium. The lensing and XSZ reconstructions are based on synthetic observation that contain observational noise. For such relaxed clusters both the lensing and the XSZ reconstructions agree very well with the original mass profile.}
\label{fig:simulated_relaxed}
\end{figure*}
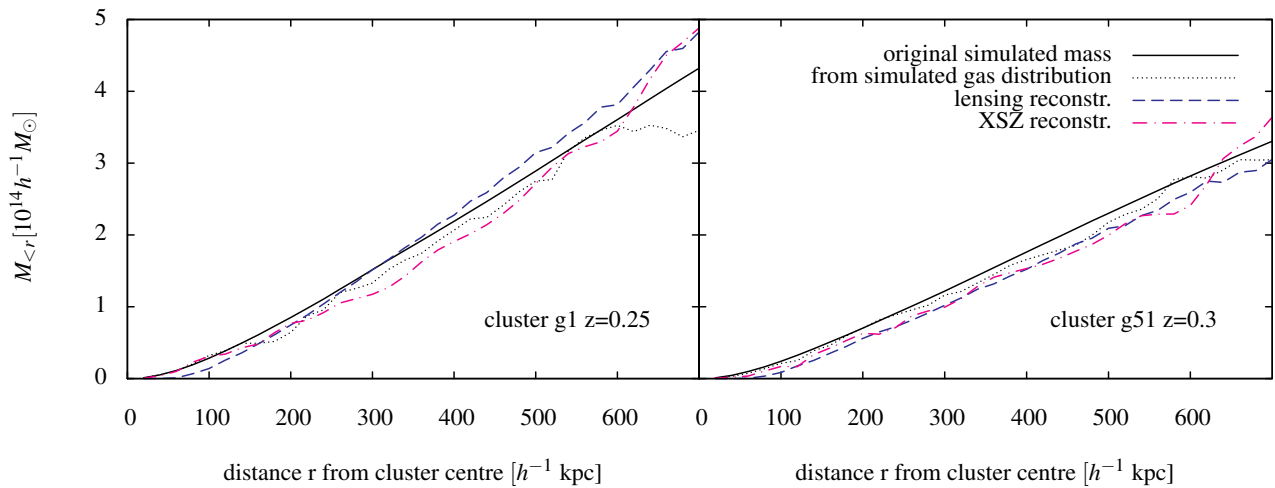 

It is reassuring that this novel method to probe hydrostatic equilibrium works well for clusters that do not have a record of recent mergers. However clusters that do experience such violent events may be even more interesting to study. In Figure \ref{fig:merger} we show a cluster at four different times during a merger. For each of these snapshots we show reconstructions of the cumulative mass profile from synthetic X-ray, SZ and lensing observations, as well as the original mass profile and the profile obtained from the simulated gas distribution. Again observational noise was added to the synthetic maps used for the reconstruction. We also show X-ray maps of the cluster for each of the four snapshots. These are however idealised noise-free versions and just meant to illustrate what is going on in the cluster. To facilitate following the merger we also show the approximate trajectory of the relevant infalling subhalo in the X-ray maps. 

In the first snapshot (upper left panel) the main cluster halo is still close to hydrostatic equilibrium. The lensing and XSZ mass estimates still agree well for radii $r$ smaller than the distance to the infalling subhalo. In the second snapshot (upper right panel), after the subhalo has passed the main halo, shocked gas causes a too large XSZ mass estimate from roughly the subhalo distance outwards. The mass profile obtained directly from the simulated gas distribution shows the same behaviour and thus confirms that this is not an artefact of the reconstruction but a real, significant deviation from hydrostatic equilibrium, which is recovered by the reconstruction or in this example even somewhat overestimated. The lensing reconstruction still reproduces the original simulated mass profile well. Thus by comparing lensing and XSZ cumulative mass profiles one can directly see the deviations from hydrostatic equilibrium. The third snapshot (lower left panel) shows that when the bow shock moves outward one can also obtain too low cluster masses by assuming hydrostatic equilibrium during a merger. Again the effect can be seen in both the mass profiles obtained directly from the simulated gas distribution and obtained from the three-dimensional XSZ gas reconstruction. In the fourth snapshot (lower right panel) hydrostatic equilibrium is already largely restored, even if one can still see the pronounced bow shock in the X-ray map. 

These simulations show that deviations from hydrostatic equilibrium during mergers can be faithfully recovered by the cluster reconstructions methods introduced above.

\begin{figure*}
{
\ifpdf
  \scalebox{0.40}{\includegraphics{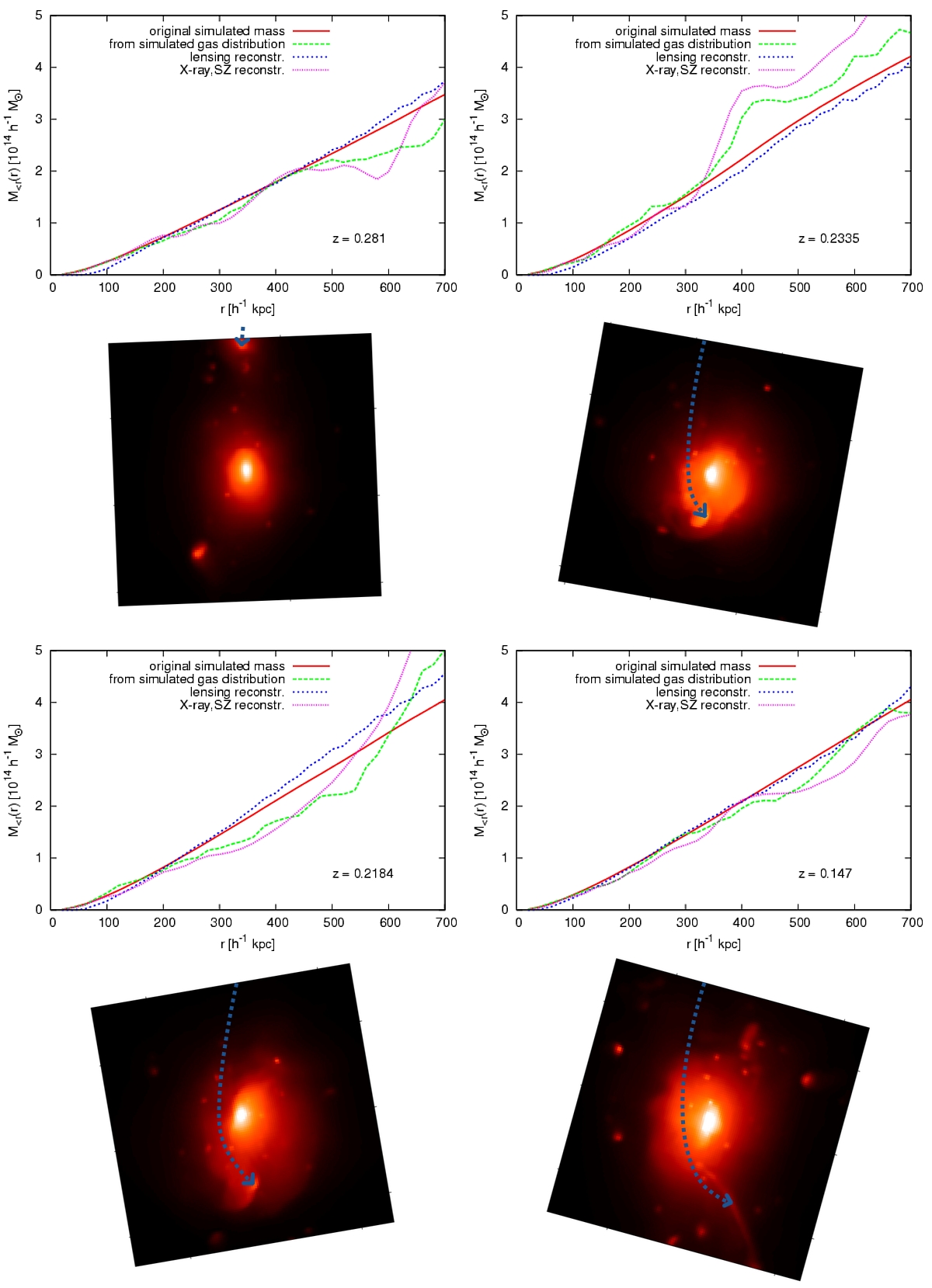}}%
\else
  \scalebox{0.82}{\includegraphics{merger}}%
\fi
}
\caption{Cumulative mass profiles $M_{<r}(r)$ and X-ray surface brightness maps of simulated cluster g51at four different redshifts during a merger. The approximate trajectory of the infalling subhalo is illustrated in the X-ray maps. Profiles of  the original simulated mass distribution, of the lensing and of the XSZ reconstructions are shown, as well as the profile obtained directly from the simulated gas distribution by assuming hydrostatic equilibrium. The lensing and XSZ reconstructions are based on synthetic observation that contain observational noise. The X-ray maps shown above are however idealised noise-free versions and were rotated such as to all have the same orientation in space.}
\label{fig:merger}
\end{figure*}

\subsection{Accuracy and reliability of cumulative mass profile reconstructions}

To determine the typical scatter in cumulative mass profile reconstructions and quantify the significance of detections of deviations from hydrostatic equilibrium we repeated the reconstruction of the merging simulated cluster shown in the upper right panel of Fig. \ref{fig:merger} with different noise realisations and for different lines-of-sight.

For the left panel of Fig. \ref{fig:mean_profiles} we used the same line-of-sight as in Fig. \ref{fig:merger} but 50 different noise realisations for the synthetic X-ray, thermal SZ and lensing observations. The noise realisations were obtained using different seeds for the random number generator employed for adding noise to the synthetic observations. The mean XSZ and lensing reconstructed profiles and the 1-$\sigma$ errors are shown as well as the profile of  the original simulated mass distribution and the profile obtained directly from the simulated gas distribution by assuming hydrostatic equilibrium. The deviations from hydrostatic equilibrium are reliably detected. As expected for a cluster that contains substructure that violates axial symmetry there are also some systematic deviations  such that the mean profiles are not centred exactly on the simulated profiles. 

For the right panel we started with a sample of synthetic lensing, X-ray and SZ observations along 50 different randomly oriented lines-of-sight. All contain realistic observational noise. It turned out that for projections for which the merging subhalo responsible for perturbing hydrostatic equilibrium is almost directly in front of or behind the main halo detecting deviations from hydrostatic equilibrium is less reliable. This is not surprising as the signal from the region where hydrostatic equilibrium is strongly perturbed is superimposed with a larger signal from the main halo, so that the contributions to such projections are difficult to separate. For the right panel of Fig. \ref{fig:mean_profiles} we thus decided to reject all 16 lines-of-sight for which the projected distance of the relevant subhalo from the main halo centre is less than $200\,h^{-1}$ kpc, as well as one line-of-sight which happened to be inclined by only $2^\circ$ with respect to the cluster's symmetry axis which is to small for a faithful reconstruction. The mean and the 1-$\sigma$ errors of the reconstructions that were based on the 33 remaining lines-of-sight are shown. Again deviations from hydrostatic equilibrium can be reliably detected. 

As discussed in Paper I reconstruction artefacts can appear close to the perimeter of the box used for the reconstructions. As we can see in the right panel of Fig. \ref{fig:mean_profiles} they can dominate the XSZ reconstructed cumulative mass profiles errors from roughly $r=600 h^{-1}$ kpc outwards for some lines-of-sight, when using a $1.5 h^{-1}$ Mpc sidelength box for the XSZ reconstruction. Thus when the quality of the observations allows studying a larger region one should also use an appropriately larger box for the XSZ reconstruction to avoid this problem.

\begin{figure*}
{
\begin{picture}(0,0)%
\scalebox{1.4}{\includegraphics{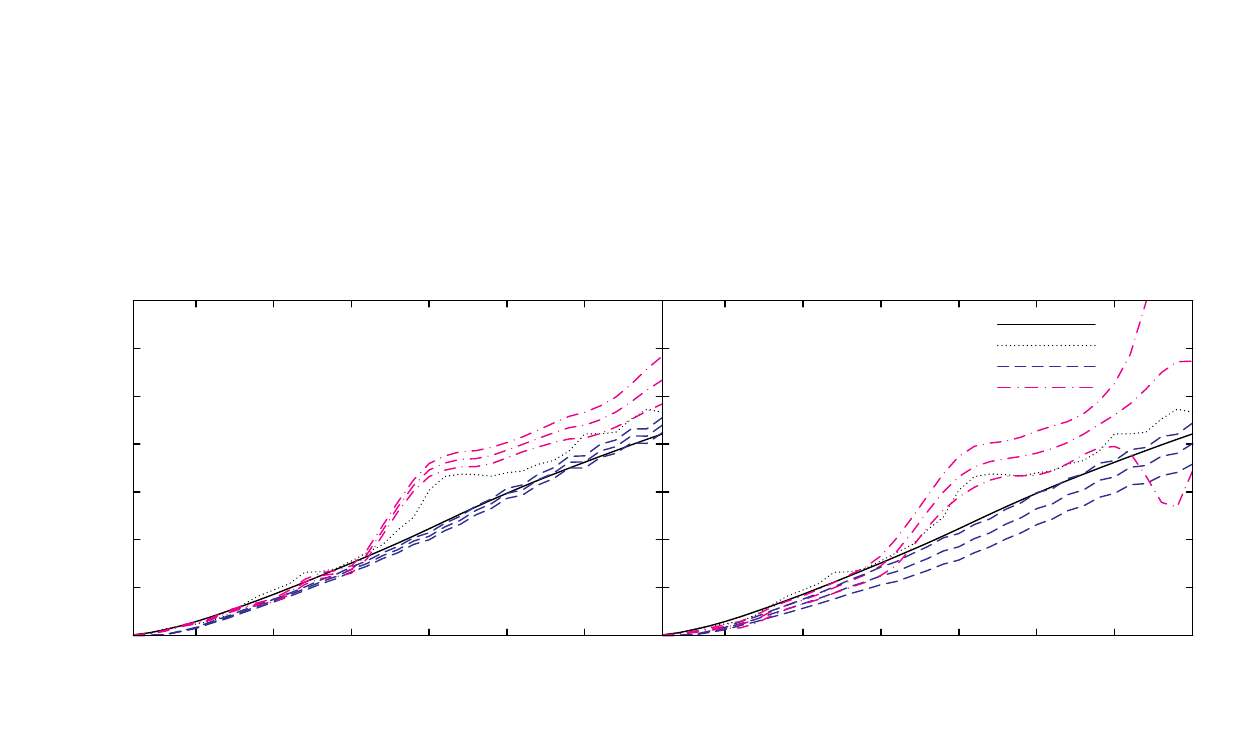}}%
\end{picture}%
\begingroup
\setlength{\unitlength}{0.0280bp}%
\begin{picture}(8459,6819)(0,0)%
\put(1650,1650){\makebox(0,0)[r]{\strut{} 0}}%
\put(1650,2339){\makebox(0,0)[r]{\strut{} 1}}%
\put(1650,3027){\makebox(0,0)[r]{\strut{} 2}}%
\put(1650,3716){\makebox(0,0)[r]{\strut{} 3}}%
\put(1650,4404){\makebox(0,0)[r]{\strut{} 4}}%
\put(1650,5093){\makebox(0,0)[r]{\strut{} 5}}%
\put(1650,5781){\makebox(0,0)[r]{\strut{} 6}}%
\put(1650,6470){\makebox(0,0)[r]{\strut{} 7}}%
\put(2821,1100){\makebox(0,0){\strut{} 100}}%
\put(3941,1100){\makebox(0,0){\strut{} 200}}%
\put(5061,1100){\makebox(0,0){\strut{} 300}}%
\put(6180,1100){\makebox(0,0){\strut{} 400}}%
\put(7300,1100){\makebox(0,0){\strut{} 500}}%
\put(8420,1100){\makebox(0,0){\strut{} 600}}%
\put(550,4060){\rotatebox{90}{\makebox(0,0){\strut{}$M_{<r} [10^{14} h^{-1} M_{\odot}]$}}}%
\put(5732,375){\makebox(0,0){\strut{}distance r from cluster centre [$h^{-1}$ kpc]}}%
\put(9265,1650){\makebox(0,0)[r]{\strut{}}}%
\put(9265,2339){\makebox(0,0)[r]{\strut{}}}%
\put(9265,3027){\makebox(0,0)[r]{\strut{}}}%
\put(9265,3716){\makebox(0,0)[r]{\strut{}}}%
\put(9265,4404){\makebox(0,0)[r]{\strut{}}}%
\put(9265,5093){\makebox(0,0)[r]{\strut{}}}%
\put(9265,5781){\makebox(0,0)[r]{\strut{}}}%
\put(9265,6470){\makebox(0,0)[r]{\strut{}}}%
\put(10438,1100){\makebox(0,0){\strut{} 100}}%
\put(11561,1100){\makebox(0,0){\strut{} 200}}%
\put(12684,1100){\makebox(0,0){\strut{} 300}}%
\put(13807,1100){\makebox(0,0){\strut{} 400}}%
\put(14929,1100){\makebox(0,0){\strut{} 500}}%
\put(16052,1100){\makebox(0,0){\strut{} 600}}%
\put(13357,375){\makebox(0,0){\strut{}distance r from cluster centre [$h^{-1}$ kpc]}}%
\put(14093,6126){\makebox(0,0)[r]{\strut{}original simulated mass}}%
\put(14093,5824){\makebox(0,0)[r]{\strut{}from simulated gas distribution}}%
\put(14093,5522){\makebox(0,0)[r]{\strut{}lensing reconstr.}}%
\put(14093,5220){\makebox(0,0)[r]{\strut{}XSZ reconstr.}}%
\put(5576,2321){\makebox(0,0)[l]{\strut{}different noise realisations}}%
\put(13803,2321){\makebox(0,0)[l]{\strut{}different lines-of-sight}}%
\end{picture}%
\endgroup
}
\caption{Mean XSZ and lensing reconstructed cumulative mass profiles $M_{<r}(r)$ and their 1-$\sigma$ errors of merging simulated cluster g51 obtained for different noise realisations (left panel) and different lines-of-sight (right panel).  Profiles of  the original simulated mass distribution and the profiles obtained directly from the simulated gas distribution by assuming hydrostatic equilibrium are shown for reference. X-ray, SZ and lensing observations along one line-of-sight but with 50 different noise realisations were used for the reconstructions whose mean and 1-$\sigma$ errors are shown in the left panel. For the right panel we started with a sample of noisy synthetic observations along 50 different randomly oriented lines-of-sight. However we rejected projections for which the merging subhalo is almost directly behind or in front of the main halo (projected distance $< 200 \, h^{-1}$ kpc) as well as one line-of-sight with an inclination of only $2^\circ$ with respect to the clusters symmetry axis which is to small for a faithful reconstruction. The mean and the 1-$\sigma$ errors of the profiles reconstructed from the observations along the remaining 33 lines-of-sight are shown. For both the different noise realisations and the different lines-of-sight deviations from hydrostatic equilibrium can be reliably detected.}
\label{fig:mean_profiles}
\end{figure*}

\section{Summary and discussion}

We proposed a novel method to obtain three-dimensional reconstructions of a galaxy cluster's gravitational potential and cumulative mass profile from X-ray and thermal SZ observations under the assumption of hydrostatic equilibrium and independently dropping this assumption from lensing data. If only X-ray and thermal SZ data is available accurate reconstructions of relaxed clusters can be obtained. If, however, lensing data is available as well, hydrostatic equilibrium can be probed, also in dynamically active clusters, by comparing these independent reconstructions.

The three-dimensional reconstructions are based on iterative Richardson-Lucy deconvolution and assume only axial symmetry of the cluster halo with respect to an arbitrarily inclined axis. The X-ray and thermal SZ data are used to first reconstruct the three-dimensional cluster gas density and temperature distribution. No equilibrium assumption except local thermal equilibrium is needed for that. Then the gravitational potential and the cumulative mass profile can be obtained from these reconstructions under the assumption of hydrostatic equilibrium. For the lensing reconstructions we deproject the lensing potential obtained by a weak lensing or a combined weak and strong lensing analysis. This yields the three-dimensional gravitational potential, from which we can get independent cumulative mass profiles by exploiting Gauss's law. The X-ray and thermal SZ analysis (abbreviated by XSZ throughout this work) and the lensing analysis are then compared in order to probe hydrostatic equilibrium and to test the accuracy of mass estimates based on the assumption of hydrostatic equilibrium.

These methods were tested with synthetic X-ray, thermal SZ and lensing observations of analytically modelled and numerically simulated galaxy clusters. Except where specifically noted realistic observational noise was added to the synthetic observations.

For analytically modelled clusters in hydrostatic equilibrium we found:

\begin{itemize}

\item Consistent and accurate lensing and X-ray, SZ based cumulative mass profiles $M_{<r,\,\mathrm{lensing}}(r)$ and $M_{<r,\,\mathrm{XSZ}}(r)$ can be obtained.

\item The accuracy somewhat decreases for very small inclination angle between the line-of-sight and the cluster's symmetry axis.

\item Higher accuracy for the iterative deprojection of the lensing potential for small inclination angles are achieved with spherically symmetrised priors. 

\item Faithful three-dimensional reconstructions of the gravitational potential can be obtained from both lensing observations and from an XSZ analysis. 

\end{itemize}

For analytically modelled clusters that are not in hydrostatic equilibrium, we showed that the deviations from equilibrium can be effectively probed by a comparison of lensing and XSZ reconstructions even when realistic observational noise is present. 
 
From reconstructions based on synthetic observations of a sample of numerically simulated galaxy clusters we conclude:

\begin{itemize}

\item For clusters that did not experience recent mergers consistent and accurate lensing and XSZ cumulative mass profiles are found. 

\item Although these clusters are not perfectly axisymmetric and noise is added to the synthetic data, the accuracy of reconstructed cumulative mass profiles is typically better than 10 to 15\% for both the X-ray, SZ and the lensing reconstructions.

\item On the other hand in clusters in the process of merging deviations from hydrostatic equilibrium can be accurately probed, except for cases where the relevant merging subhalo appears directly in front of or behind the main halo's centre. 

\end{itemize}

\acknowledgements{We are deeply indebted to Klaus Dolag, who generously provided us access to the numerical simulations of the cluster sample that was used in this work. E.~P.~was supported by the German Science Foundation under grant number BA~1369/6-1within the framework program SPP 1177.}

\bibliographystyle{aa}

\end{document}